# Analyzing Zone Routing Protocol in MANET Applying Authentic Parameter

Kamaljit I. Lakhtaria
MCA Department,
Atmiya Institute of Technology & Science
Yogidham, Rajkot, Gujarat, INDIA
Email: kamaljit.ilakhtaria@gmail.com

**Abstract**

*Routing is the main part of wireless adhoc network conventionally there are two approaches first one is Proactive and another one is Reactive. Both these approaches have some substantial disadvantage and to overcome hybrid routing protocols designed. ZRP (Zone Routing Protocol) is one of the hybrid routing protocols, it takes advantage of proactive approach by providing reliability within the scalable zone, and for beyond the scalable zone it looks for the reactive approach. It (ZRP) uses the proactive and the reactive routing according to the need of the application at that particular instance of time depending upon the prevailing scenario. This work revolves around the performance of ZRP against realistic parameters by varying various attributes such as Zone Radius of ZRP in different node density. Results vary as we change the node density on Qualnet 4.0 network simulator.*

## 1. Introduction

Mobile ad hoc networks (MANETs) [1] are collections of mobile nodes, dynamically forming a temporary network without pre-existing network infrastructure or centralized administration. These nodes can be arbitrarily located and are free to move randomly at any given time, thus allowing network topology and interconnections between nodes to change rapidly and unpredictably. MANET is likely to be use in many practical applications, including personal area networks, home area networking, and military environments, and so on recent advances in wireless technology have enhanced the feasibility and functionality of wireless mobile ad hoc networks (MANETs). There has been significant research activity over the past 10 years into performance of such networks with the view to develop more efficient and robust routing protocols. However, there is majority research has concentrated on proactive or reactive routing protocol for data transmission, improving performance metrics and on the Security threats of this protocol by making change in it. But proactive and reactive both have some disadvantage as hybrid routing protocol come into existence is combination of both proactive and reactive, ZRP one among them come in to existence. Our contributions are as follows: Section I, introduces ZRP protocol and its component Section II, give details of previous and related work. In section III, we discuss about the simulation environment, in section IV, we discuss the result and in Section V, we conclude all the work and future work.

### I. 1 ZRP (ZONE ROUTING PROTOCOL)

ZRP [6] is a framework by using it we can take advantage of both table driven and on demand driven protocol according to the application. In this separation of nodes, local neighborhood from the global topology of the entire network allows for applying different approaches and thus taking advantage of each technique's features for a given situation. These local neighborhoods are called *zones* (hence the name) each node may be within multiple overlapping zones, and each zone may be of a different size. The "size" of a zone is not determined by geographical measurement, as one might expect, but is given by a radius of length **α** where **α** is the number of hops to the perimeter of the zone.

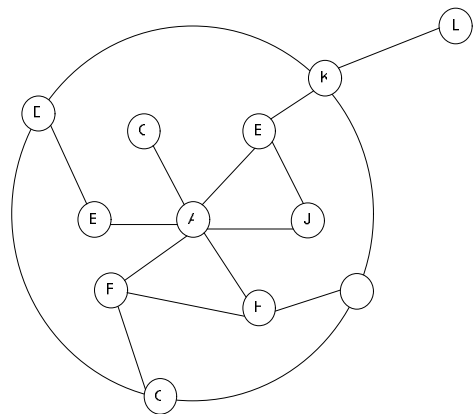

Figure 1 ZRP having Zone radius α =2

In the above diagram ZRP, protocol having Zone radius 2 in this in side the zone communication done in proactive way and out side it between such zones in reactive way. A, E, F, H, J, C are interior node and D, G, I, k are border nodes communication between B and K is done through proactive way and L is located out side the zone. ZRP consist of [8] three parts IARP [9] proactive part,

IERP [10] reactive part of it and BRP [11] used with IERP to reduce the query traffic.

### I.2 IARP (INTRA ZONE ROUTING PROTOCOL)

The Intra zone Routing Protocol (IARP) [9] is a limited scope proactive routing protocol, which used to support a primary global routing protocol. The routing zone radius shows the scope of the proactive part, the distance in hops that IARP route updates relayed. IARP's proactive tracking of local network connectivity provides support for route acquiring and route maintenance. First, routes to local nodes are immediately available, avoiding the traffic overhead and latency of a route discovery. Traditional proactive link state protocols modified to serve as an IARP by limiting link state updates to the scope of the link source's routing zone.

### I.3 IERP (INTER ZONE ROUTING PROTOCOL)

The Interzone Routing Protocol (IERP) is the global reactive routing component of the Zone Routing Protocol (ZRP)[6].IERP adapts existing reactive routing protocol implementations to take advantage of the known topology of each node's surrounding R-hop neighborhood (routing zone), provided by the Interzone outing Protocol (IARP)[9]. The availability of routing zone routes allows IERP to suppress route queries for local destinations. When a global route discovery is required, the routing zone based border cast service [11] used for efficiently guide route queries outward, rather than blindly relaying queries from neighbor to neighbor. Once a route discovered, IERP can use routing zones automatically to redirect data around failed links similarly, suboptimal route segments identified and traffic re-routed along shorter paths.

### I.4 BRP (BORDERCAST RESOLUTION PROTOCOL)

The Bordercast Resolution Protocol (BRP) [11] provides the bordercasting packet delivery service. The BRP uses a map of an extended routing zone, provided by the local proactive Intrazone Routing Protocol (IARP) [9], to construct Bordercast (multicast) trees along which query packets are directed. (Within the context of the hybrid ZRP, the BRP used to guide the route requests of the global reactive Interzone Routing Protocol (IERP) [10]). The BRP uses special query control mechanisms to steer route requests away from areas of the network that have already covered by the query. The combination of multicasting and zone based query control makes Bordercasting an efficient and tunable service that is more suitable than flood searching for network probing applications like route discovery. The Bordercast Resolution Protocol (BRP) is a packet delivery service, not a full featured routing protocol. Bordercasting enabled by local proactive Intrazone Routing Protocol (IARP) and supports global reactive Interzone Routing Protocol (IERP).

## II RELATED WORKS

Nicklas Beijar in 2001[5] first discuss the problem in proactive and reactive routing and then how they move towards the ZRP (Zone Routing Protocol) paper describe the architecture of the ZRP also describe the working of the protocol with an example. In 2002 Jan Schaumann [6] analyze the ZRP in mobile Adhoc network discuss the basic of MANET and implication on routing and problems occur due to rapidly changing topology without fixed router. In paper author, also discuss the ZRP hybrid routing protocol having both proactive and reactive protocol in context to other routing protocol. In 2003, David Oliver Jorg discusses the performance comparison of MANET routing protocol in different network size in that paper they discuss the problem due to the mobility of different nodes they test the routing performance of four different routing protocol. [7]in this examine the analytical simulation result for the routing protocol DSR ,TORA and ZRP emphasizing on the ZRP and impact of some of it most important attributes to the network performance. Julian Hsu, Sameer Bhatia, Mineo Takai, Rajive Bagrodia,[13] discuss the performance of common MANET routing protocol under realistic scenarios protocols include AODV OSPFv2 and ZRP which comprise all proactive, reactive ,hybrid routing protocol. In [14] discuss some of the factor that affects the routing algorithm like such as variable wireless link quality, propagation path loss, fading; multi-user interference, power expended and topological changes become important issues.. In paper, discuss about the proactive DSDV, WRP, CGSR, reactive SSR, AODV, RDMAR, Hybrid routing protocol like, ZRP. In [15] paper presents the idea of integrating the layer-II label-switching technique with layer-III and study the effect of MultiProtocol Label Switch (MPLS) mechanism on the performance Ad-Hoc Networks (MANETs). In 2007[16] discuss the performance of three routing protocol DSR, AODV, LAR1 the performance is analyzed using varying, mobility and network size perform simulation on GLOMOSIM network simulator.

## III SIMULATION ENVIRONMENT

The simulation work done on Qualnet wireless network simulator version 4.0. Mobility model used is Random Way Point (RWP). In this model a Mobile node is initially placed in a random location in the simulation area, and then moved in an anomaly chosen direction between [0, 2] at a random speed between [SpeedMin, SpeedMax]. The movement proceeds for a specific amount of time or distance, and the process is repeated a predetermined number of times. We chose Min speed = 0 m/s, Max speed = 10m/s, and pause time = vary. All the simulation work was carried out using ZRP routing protocol. Network traffic is provided by using Constant Bit Rate (CBR) sources. A CBR traffic source provides a constant stream of packets throughout the whole simulation thus further stressing the routing task.

### III.1 Parameter Value For Simulation

- Mobility model Random Wave Point
  Minimum speed 0 mps
  Maximum speed 10 mps
- Pause time 30s
- Simulation Time 120s
  Terrain
  Coordination 800 * 800 m
    Connection
  CBR (Constant Bit Ratio)
  Item size 512(byte)
    Radio/physical layer parameters
  Radio type: 802.11b
  Data rate: 2Mbps

### III .2 Efficiency Metrics Used

Throughput: It is the measure of the number of packets successfully transmitted to their final destination per unit time. It is the ratio between the numbers of sent packets vs. received packets.

Avg End to END Delay: It signifies the average time taken by packets to reach one end to another end (Source to Destination).

Avg Jitter Effect: It signifies the Packets from the source will reach the destination with different delays. A packet's delay varies with its position in the queues of the routers along the path between source and destination and this position can vary unpredictably.

Packet Loss Percentage: It is the Ratio of transmitted packets that may have been discarded or lost in the network to the total number of packet sent.

## IV RESULTS

Figure2 depicted that throughput of the ZRP having smaller zone radius decreases as compared to ZRP having higher zone radius as the node density increases. The possible reasons are as node density increases number of neighbor around the node increases and number of zones in the area increases. Due to this number of zones increases, so that reactive traffic of ZRP increases as compared to proactive one and large number of query packet are generated, to share information between zones.

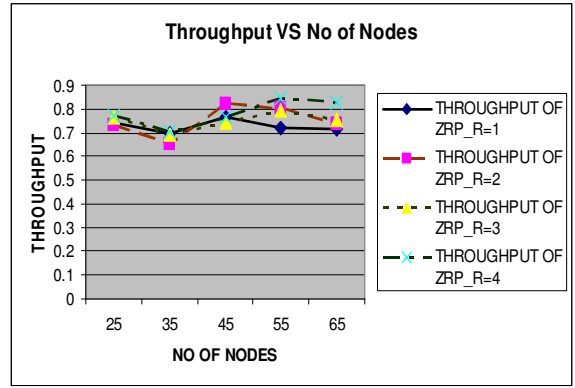

Figure 2 Comparison of throughput of ZRP in different node density by varying Zone radius.

Hence, large numbers of query packets are generated so chances of wrong path selection and time required for searching the destination increases.

However, on the other side throughput of ZRP, having higher zone radius is better then the ZRP having smaller zone radius as the node density increases. The possible reasons are as the zone radius is increased zone size also increases and proactive traffic in ZRP increases as compared to reactive. Therefore, nodes have details of large number of neighbor around them, chance of query packet, data packet loss is less, and time required to share information with global part is decreases. As above, all discussion shows that ZRP having higher zone radius give the better throughput as compared to ZRP having smaller zone radius in high-density nodes.

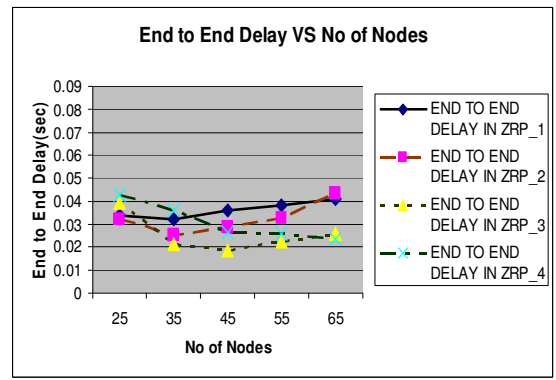

Figure 3 Comparison of End to End delay of ZRP in different node density by varying Zone radius.

Figure3 depicted that when the node density is less, ZRP having all zone radius almost give same end-to-end delay. The possible reason for this is as the density of node is less, number of neighbor around the node is less. Therefore, less number of update messages is required to take the details of nodes and time required to share information to exterior part reduced. Hence, overall delay required by the packet to reach destination from the source is almost same for all zone radius in ZRP.

However on the other side when node density increases end-to-end delay increases, in the ZRP having smaller zone radius as compared to ZRP having higher zone radius. The possible reason for this is as zone radius is smaller, number of zone increases. Due to this reactive traffic increases and chance of query, packet loss is also more and time required to share information between zone increases. Therefore, due to all these overall time delay required by the packet to reach the destination form the source increases. On the other hand, ZRP having higher zone radius shows less end-to-end delay as compared to ZRP having smaller zone radius. The possible reason is as the zone radius increases zone size also increases and proactive traffic of the ZRP used more as compared to reactive. Hence details of large number of node is available, so less time is required to share the information with global part, because of all this over all time delay taken by the packet to reach destination form source is reduced. Above all discussion shows that ZRP having higher zone radius produce less end-to-end delay as compared to ZRP having smaller zone radius in high-density node.

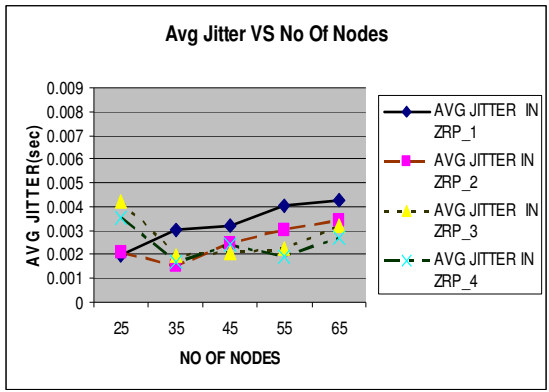

Figure 4 Comparison of Avg jitter Effect of ZRP in different node density by varying Zone Radius.

Figure 4 depicted that Avg jitter effect increases in ZRP having smaller zone radius as compared to ZRP having higher zone radius when node density increases. The possible reason for this is as zone radius is small number of zone increases, and reactive traffic in the ZRP increases as compared to proactive. Therefore, large numbers of query packet generated to search the path between zones. In these chances of query packet loss increases, hence time required for sharing information between zones vary, because of this packet form source reach the destination at different time delay.

However, on the other side Avg jitter effect is less in ZRP having higher zone radius as compared to ZRP having smaller zone radius in high-density node. The possible explanation is as the zone radius increases zone size also increases and number of zone reduced. Due to this proactive traffic in ZRP is more as compared to reactive traffic. Therefore, a detail of large number of nodes is available so chances of query packet loss are less. Due to this time required sharing, information with global part reduced and packet form the source to destination reach at equal interval.

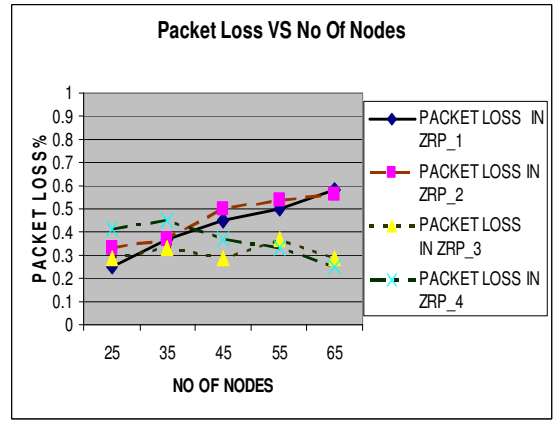

Figure 5 Comparison of packet loss using ZRP in different node density by varying Zone Radius.

Figure 5 depicted that when node density increases packet loss increases in ZRP smaller zone radius, as compared ZRP having higher zone radius. The possible explanation is as the node density high number of neighbor around the node increases, and number of zone increases. Due to this reactive traffic in ZRP, is more as compared to proactive. Therefore, a chance of query packet, data packet loss and wrong path selection increases.

However, on the other side as node density increases packet loss is less in ZRP having high zone radius as compared to ZRP having smaller zone radius. The possible explanation is as the node density high number of neighbor around the node also increases. Moreover, if zone radius is higher zone size increases, and number of zones decreases. Hence, proactive traffic in ZRP is more as compared to reactive and zone size is large so details of larger number of nodes are available. The reactive part is less, chances of query packet loss and packet loss due to wrong path selection also reduced.

## V. CONCLUSIONS AND FUTURE WORK

Node density has truly shown the effect on the performance of the ZRP protocol. As the density, changes ZRP attribute Zone radius has to be changed to get good performance. Result shows that configuration of Zone radius according to what type of application in which we use ZRP protocol. The high-density increases may increase the discovered services but it deteriorates there quality in terms of availability. If it is used for real time application likes video transmission then due to jitter effect performance decreases. In other application in which delay is consider then we can use the reduced Zone radius. Because as we increase the proactive part by increasing the Zone radius control traffic also increases. ZRP is suitable for the large network by providing the benefit of both proactive and reactive routing protocol.

As part of our future work we simulate ZRP by varying mobility and check its performance. Also check the performance of ZRP without using BRP it is interesting to see the performance of ZRP in large and realistic scenario.